\documentclass[prd,aps,reprint,onecolumn]{revtex4-1}

\usepackage{hyperref,amsfonts,amsmath,amssymb,graphicx,bm,subfigure}
\usepackage{color}
\usepackage{soul}

\begin{document}

\author{$^{a}$\,Mridupawan Deka}
\email{mpdeka@theor.jinr.ru}

\author{$^{a,b}$\,Maxim Dvornikov}
\email{maxdvo@izmiran.ru}

\title{Spin oscillations in neutrino gravitational scattering}

\affiliation{$^{a}$\,Bogoliubov Laboratory of Theoretical Physics, Joint Institute for Nuclear Research, 141980 Dubna, Moscow region, Russia; \\
$^{b}$\,Pushkov Institute of Terrestrial Magnetism, Ionosphere
and Radiowave Propagation (IZMIRAN),
108840 Moscow, Troitsk, Russia}

\begin{abstract}
We study neutrino spin oscillations while the particles scatter off a rotating black hole surrounded by a thick magnetized accretion disk. Neutrino spin precession is caused by the interaction of the neutrino magnetic moment with the magnetic field in the disk which has both toroidal and poloidal components. Our calculation of the observed neutrino fluxes, accounting for spin oscillations, are based on numerical simulations of the  propagation of a great number of incoming test particles using High Performance Parallel Computing. The obtained results significantly improve our previous findings. We briefly discuss the applications for the observations of astrophysical neutrinos.
\end{abstract}

\maketitle

\section{Introduction}
\label{sec:INTR}

Nonzero masses of neutrinos as well as the mixing between different neutrino generations result in neutrino flavor oscillations~\cite{GiuKim07}, i.e. the flavor content of a neutrino beam will change while this beam propagates in space. The nontrivial electromagnetic properties of neutrinos~\cite{Giu16}, namely, the presence of neutrino magnetic moments, lead to neutrino spin and spin-flavor oscillations. In this situation, left polarized active neutrinos are converted to sterile right handed particles under the influence of external magnetic fields.

The gravitational interaction is known to affect the motion and spin evolution of a fermion (see, e.g., Ref.~\cite{Ver22}). In general case, the motion of a spinning particle deviates from the geodesics~\cite{Wal72}. However, in case of a point-like elementary particle, i.e., a neutrino, this deviation is negligible. The quasiclassical approach for the study of the spinning particles motion in a curved spacetime was put forward in Ref.~\cite{PomKhr98}. This approach has been adapted in Ref.~\cite{Dvo06} for the description of neutrino spin oscillations. The contributions of the neutrino electromagnetic and electroweak interactions have been incorporated in this approach in Ref.~\cite{Dvo13}.

Spin oscillations of the scattered neutrinos off a black hole (BH) have been studied in Refs.~\cite{Dvo23c,Dvo23d,Dvo23a,Dvo23b}. The particle gravitational scattering is the most appropriate way to study the effect of gravity on the spin evolution since neutrino in and out states are in the asymptotically flat spacetime. The neutrino polarization is well defined in this situation.

In the present work, following Ref.~\cite{Dvo23b}, we consider a rotating supermassive BH (SMBH) surrounded by a thick magnetized accretion disk. The model of such a disk was put forward in Ref.~\cite{Kom06}. The main  improvement compared to Ref.~\cite{Dvo23b} consists in the using of a great number of incoming test neutrinos. It allows one to significantly increase the resolution of fluxes of the scattered neutrinos. However, we have to use High Performance Parallel Computing to reach our goal.

This work is organized in the following way. First, in Sec.~\ref{sec:MOTION}, we briefly review how to describe the motion and the spin evolution of a test particle in the gravitational field of a rotating BH. The interactions of the neutrino spin with background matter and magnetic fields are also discussed. The results of the calculations of the scattered neutrino fluxes are present in Sec.~\ref{sec:RES}. Finally, in Sec.~\ref{sec:CONCL}, we conclude.

\section{Motion and spin evolution of ultrarelativistic neutrinos}
\label{sec:MOTION}

The metric of spacetime of a rotating BH with the mass $M$ and the angular momentum $J$, which is along the $z$-axis, has the form,
\begin{equation}\label{eq:Kerrmetr}
  \mathrm{d}s^{2}=g_{\mu\nu}\mathrm{d}x^{\mu}\mathrm{d}x^{\nu}=
  \left(
    1-\frac{rr_{g}}{\Sigma}
  \right)
  \mathrm{d}t^{2}+2\frac{rr_{g}a\sin^{2}\theta}{\Sigma}\mathrm{d}t\mathrm{d}\phi-\frac{\Sigma}{\Delta}\mathrm{d}r^{2}-
  \Sigma\mathrm{d}\theta^{2}-\frac{\Xi}{\Sigma}\sin^{2}\theta\mathrm{d}\phi^{2},
\end{equation}
where $r_g = 2M$ is the Schwarzschild radius, $a = J/M$, and
\begin{equation}\label{eq:dsxi}
  \Delta=r^{2}-rr_{g}+a^{2},
  \quad
  \Sigma=r^{2}+a^{2}\cos^{2}\theta,
  \quad
  \Xi=
  \left(
    r^{2}+a^{2}
  \right)
  \Sigma+rr_{g}a^{2}\sin^{2}\theta.
\end{equation}
We use the Boyer-Lindquist coordinates $x^{\mu}=(t,r,\theta,\phi)$ in Eqs.~\eqref{eq:Kerrmetr} and~\eqref{eq:dsxi}.

The trajectory of ultrarelativistic neutrinos, i.e. the dependencies $\theta(r)$ and $\phi(r)$, in the gravitational field of a rotating BH can be obtained from the integral expressions~\cite{GraLupStr18},
\begin{align}
  & \phi = a\int\frac{\mathrm{d}r}{\pm\Delta\sqrt{R}}[(r^{2}+a^{2})E-aL]+
  \int\frac{\mathrm{d}\theta}{\pm\sqrt{\Theta}}
  \left[
    \frac{L}{\sin^{2}\theta}-aE
  \right],
  \label{eq:trajphi}
  \\
  & \int\frac{\mathrm{d}r}{\pm\sqrt{R}}=\int\frac{\mathrm{d}\theta}{\pm\sqrt{\Theta}},
  \label{eq:trajth}
\end{align}
where
\begin{align}\label{eq:RTh}
  R(r) = & [(r^{2}+a^{2})E-aL]^{2}-\Delta[Q+(L-aE)^{2}],
  \notag
  \\
  \Theta(\theta)= & Q + \cos^{2}\theta
  \left(
    a^{2} E^{2} - \frac{L^{2}}{\sin^{2}\theta} 
  \right).
\end{align}
In Eqs.~\eqref{eq:trajphi}-\eqref{eq:RTh}, we take into account the conserved quantities in the particle motion in the Kerr metric in Eq.~\eqref{eq:Kerrmetr}: the energy $E$, the projection of the particle angular momentum on the $z$-axis $L$, and the Carter constant $Q$. The signs $\pm{}$ in the integrands in Eqs.~\eqref{eq:trajphi} and~\eqref{eq:trajth} correspond to those of $\mathrm{d}r$ and $\mathrm{d}\theta$ in a particular branch of the trajectory.

We suppose that the incoming particles are emitted from the source with the coordinates $(r,\theta,\phi)_s = (\infty,\pi/2,0)$. The observer is in the position $(\infty,\theta_\mathrm{obs},\phi_\mathrm{obs})$, where $\theta_\mathrm{obs}$ and $\phi_\mathrm{obs}$ are arbitrary quantities. If a neutrino experiences the scattering, then $Q>0$.

The main difficulty in the analysis of Eq.~\eqref{eq:trajth} is the inversions $N$ of the trajectory with respect to the equatorial plane. If an incoming neutrino is above the equatorial plane, the number of inversions, when the particle propagates from the infinity to the turn point, is
\begin{equation}\label{eq:Ninv}
  N =
  \left\lfloor
    \frac{1}{2}
    \left(
      \frac{I_x}{K} -1
    \right)  
   \right\rfloor
   +1,
\end{equation}
where
\begin{align}
  I_x = & z \sqrt{t_+^2 + t_-^2}
  \int_x^\infty \frac{\mathrm{d}x'}{\sqrt{x'^{4}+x'^{2}\left[z^{2}-w-y^{2}\right]+x'\left[w+\left(z-y\right)^{2}\right]-z^{2}w}},
  \notag
  \\
  K = & K
  \left(
    \frac{t_{+}^{2}}{t_{-}^{2}+t_{+}^{2}}
  \right),
  \quad
  t_{\pm}^{2}=\frac{1}{2z^{2}}\left[\sqrt{(z^{2}-y^{2}-w)^{2}+4z^{2}w}\pm(z^{2}-y^{2}-w)\right].
\end{align}
Here $K(m)$ is the complete elliptic integral of the first kind, as well as $r=xr_{g}$, $L=yr_{g}E$, $Q=wr_{g}^{2}E^{2}$, and $a=zr_{g}$ are the dimensionless variables. The dependence $\theta(x)$ can be obtained from,
\begin{equation}\label{eq:thetabtp}
  \cos\theta = t_+ \text{cn}
  \left(
      (-1)^N \left\{ K \left(4 \left\lfloor \frac{N}{2} \right\rfloor +1 \right) - I_x \right\} \bigg| \frac{t_{+}^{2}}{t_{-}^{2}+t_{+}^{2}}
  \right),
\end{equation}
where $\text{cn}(u|m)$ is the elliptic Jacobi function and $N$ is given in Eq.~\eqref{eq:Ninv}. The definitions of elliptic integrals and functions correspond to those in Ref.~\cite{AbrSte64}.

If a neutrino moves from the turn point towards the infinity, then
\begin{equation}
  N =
  \begin{cases}
  \left\lfloor
    \frac{I_x + F}{2K} 
   \right\rfloor, & \text{if} \quad \dot{t}_\mathrm{pt} < 0,
  \\
  \left\lfloor
    \frac{I_x - F}{2K} 
   \right\rfloor + 1, & \text{if} \quad \dot{t}_\mathrm{pt} >0,
  \end{cases}
\end{equation}
where ${t}_\mathrm{pt} = \cos \theta_\mathrm{pt}$ corresponds to the turn point and 
\begin{equation}
  F = F
  \left(
    \arccos \left( \frac{t_\mathrm{pt}}{t_{+}} \right),\frac{t_{+}^{2}}{t_{-}^{2}+t_{+}^{2}}
  \right).
\end{equation}
Here $F(\varphi,m)$ is the incomplete elliptic integral of the first kind. The function $\theta(x)$ results from,
\begin{equation}\label{eq:thetaatp}
  \cos\theta =
  t_+\times
  \begin{cases}
    \text{cn}
    \left(
      (-1)^N
      \left(
        I_x + F - 4 K
        \left\lceil
          \frac{N}{2}
        \right\rceil
      \right)
      \Big| \frac{t_{+}^{2}}{t_{-}^{2}+t_{+}^{2}}
    \right),
    & \text{if} \quad \dot{t}_\mathrm{pt} < 0,
    \\
    \text{cn}
    \left(
      (-1)^N
      \left(
        F - I_x + 4 K
        \left\lfloor
          \frac{N}{2}
        \right\rfloor
      \right)
      \Big| \frac{t_{+}^{2}}{t_{-}^{2}+t_{+}^{2}}
    \right),
  & \text{if} \quad \dot{t}_\mathrm{pt} >0.
  \end{cases}
\end{equation}
We find in Ref.~\cite{Dvo23b} that one can describe the motion and the spin evolution of the incoming neutrinos which move below the equatorial plane using the symmetry arguments.

We suppose that, besides gravity, a neutrino interacts electroweakly with the accretion disk, which surrounds BH. Moreover, a neutrino is assumed to have a nonzero magnetic moment $\mu$ which causes the particle spin precession in electromagnetic fields arising in the accretion disk. We adopt the magnetized Polish doughnut model for the accretion disk which was developed in Ref.~\cite{Kom06}.

Besides the toroidal magnetic field, predicted in Ref.~\cite{Kom06}, we assume that a disk is permeated by a poloidal magnetic field. We consider two models of such a field proposed in Refs.~\cite{Wal74,FraMei09}. First, we study the poloidal field based on the vector potential
\begin{equation}\label{eq:Atphi}
  A_{t}=Ba
  \left[
    1-\frac{rr_{g}}{2\Sigma}(1+\cos^{2}\theta)
  \right],
  \quad
  A_{\phi}=-\frac{B}{2}
  \left[
    r^{2}+a^{2}-\frac{a^{2}rr_{g}}{\Sigma}(1+\cos^{2}\theta)
  \right]\sin^{2}\theta,
\end{equation}
where $B$ scales with distance as $B \propto B_0 r^{-5/4}$~\cite{BlaPay82}, where $B_0$ is the strength in the vicinity of the BH surface. Second, we consider the field corresponding to
\begin{equation}\label{eq:Aphi}
  A_{\phi}=b\rho,
\end{equation}
where $\rho$ is the mass density of the accretion disk and $b$ is the phenomenological constant.

The neutrino polarization is described by the invariant three vector $\bm{\zeta}$ which is defined in the rest frame with respect to the locally Minkowskian frame. The vector $\bm{\bm{\zeta}}$ obeys the evolution equation,
\begin{equation}\label{eq:nuspinrot}
  \frac{\mathrm{d}\bm{\bm{\zeta}}}{\mathrm{d}t}=2(\bm{\bm{\zeta}}\times\bm{\bm{\Omega}}),
  \quad
  \bm{\bm{\Omega}}=\bm{\bm{\Omega}}_{g}+\bm{\bm{\Omega}}_{\mathrm{em}}+\bm{\bm{\Omega}}_{\mathrm{matt}},
\end{equation}
where $\bm{\bm{\Omega}}_{g,\mathrm{em,matt}}$ are the contributions of the gravitational, electromagnetic, and the electroweak interactions of a neutrino. The explicit forms of these vectors are given in Refs.~\cite{Dvo23a,Dvo23b}.

It is more convenient to study the effective Schr\"odinger equation for the description of the neutrino polarization,
\begin{align}\label{eq:Schreq}
  \mathrm{i}\frac{\mathrm{d}\psi}{\mathrm{d}x}= & \hat{H}_{x}\psi,
  \quad
  \hat{H}_{x}=-\mathcal{U}_{2}(\bm{\bm{\sigma}}\cdot\bm{\bm{\Omega}}_{x})\mathcal{U}_{2}^{\dagger},
  \notag
  \\
  \bm{\bm{\Omega}}_{x} = & r_{g}\bm{\bm{\Omega}}\frac{\mathrm{d}t}{\mathrm{d}r},
  \quad
  \mathcal{U}_{2}=\exp(\mathrm{i}\pi\sigma_{2}/4),
\end{align}
where $\bm{\bm{\sigma}}=(\sigma_{1},\sigma_{2},\sigma_{3})$ are the Pauli matrices. The Hamiltonian $\hat{H}_{x}$ is the function of $x$ only since the dependence $\theta(x)$ is given in Eqs.~\eqref{eq:thetabtp} and~\eqref{eq:thetaatp}. The initial condition for Eq.~(\ref{eq:Schreq}) has the form, $\psi_{-\infty}^{\mathrm{T}}=(1,0)$, which corresponds to left polarized incoming neutrinos. The polarization of a scattered particle is described by the solution of Eq.~(\ref{eq:Schreq}) in the form, $\psi_{+\infty}^{\mathrm{T}}=(\psi_{+\infty}^{(\mathrm{R})},\psi_{+\infty}^{(\mathrm{L})})$.
The survival probability, i.e. the probability that a neutrino remains
left polarized after the scattering, is $P_{\mathrm{LL}}=|\psi_{+\infty}^{(\mathrm{L})}|^{2}$. 

\begin{figure}[htbp]
  \centering
  \subfigure[]
  {\label{fig:f1a}
  \includegraphics[scale=.33]{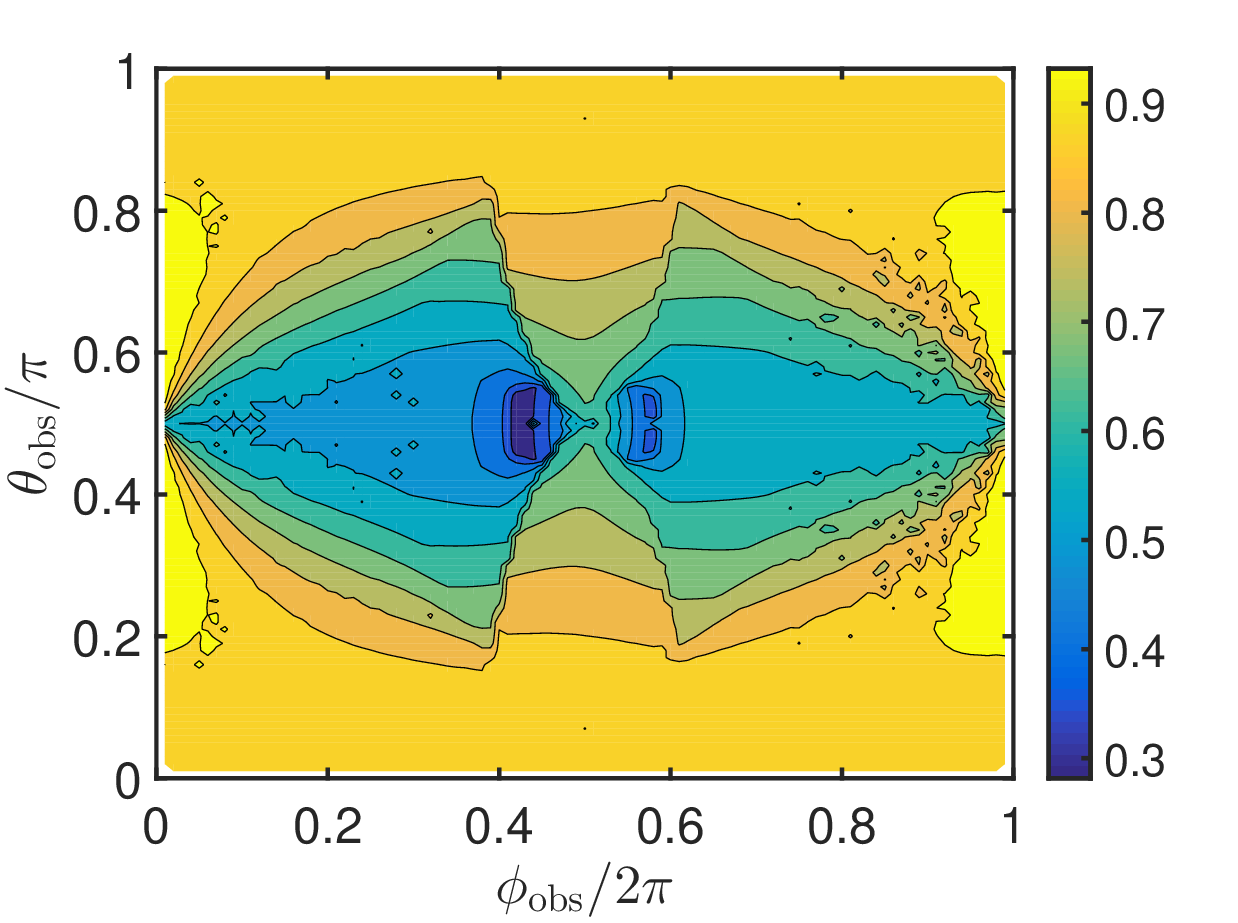}}
  \subfigure[]
  {\label{fig:f1b}
  \includegraphics[scale=.33]{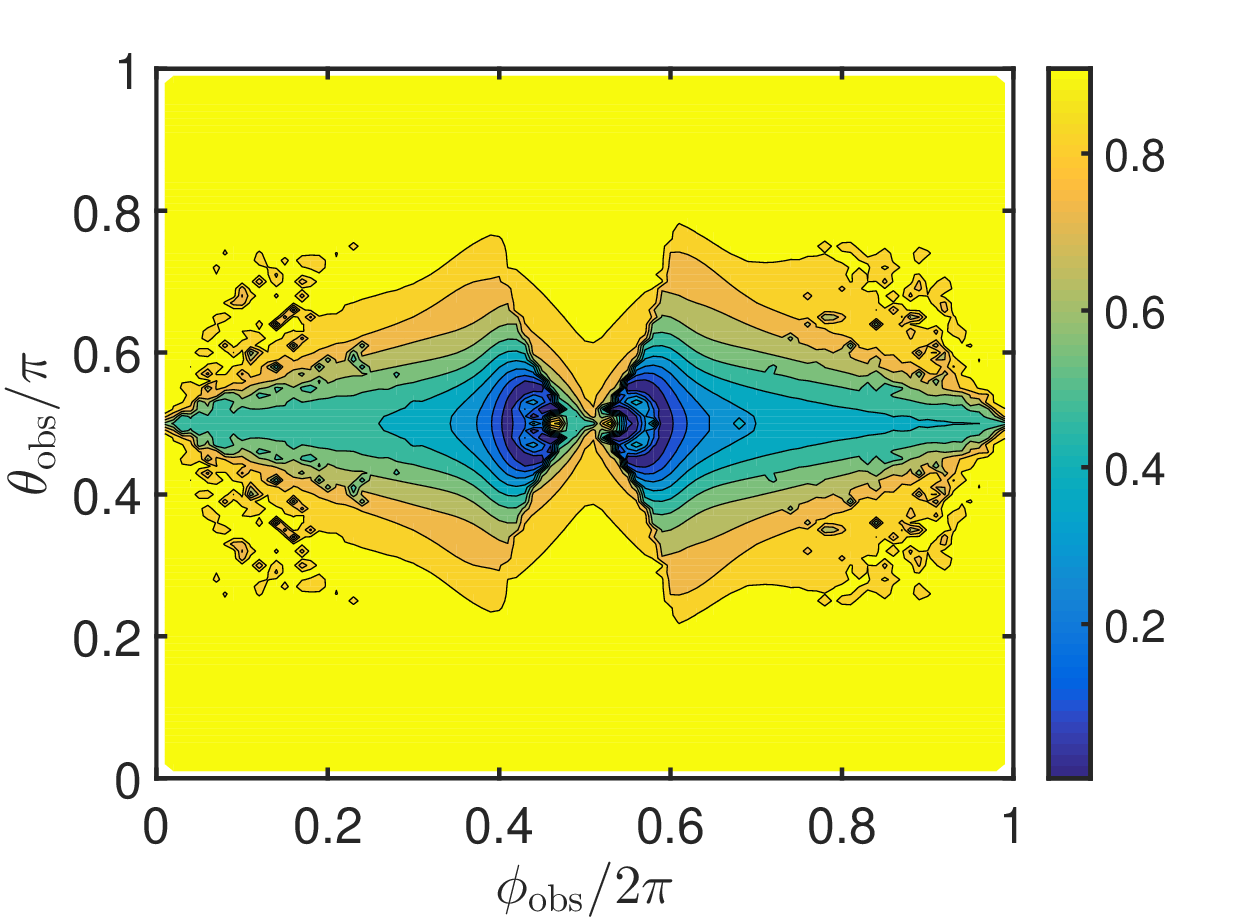}}
  \\
  \subfigure[]
  {\label{fig:f1c}
  \includegraphics[scale=.33]{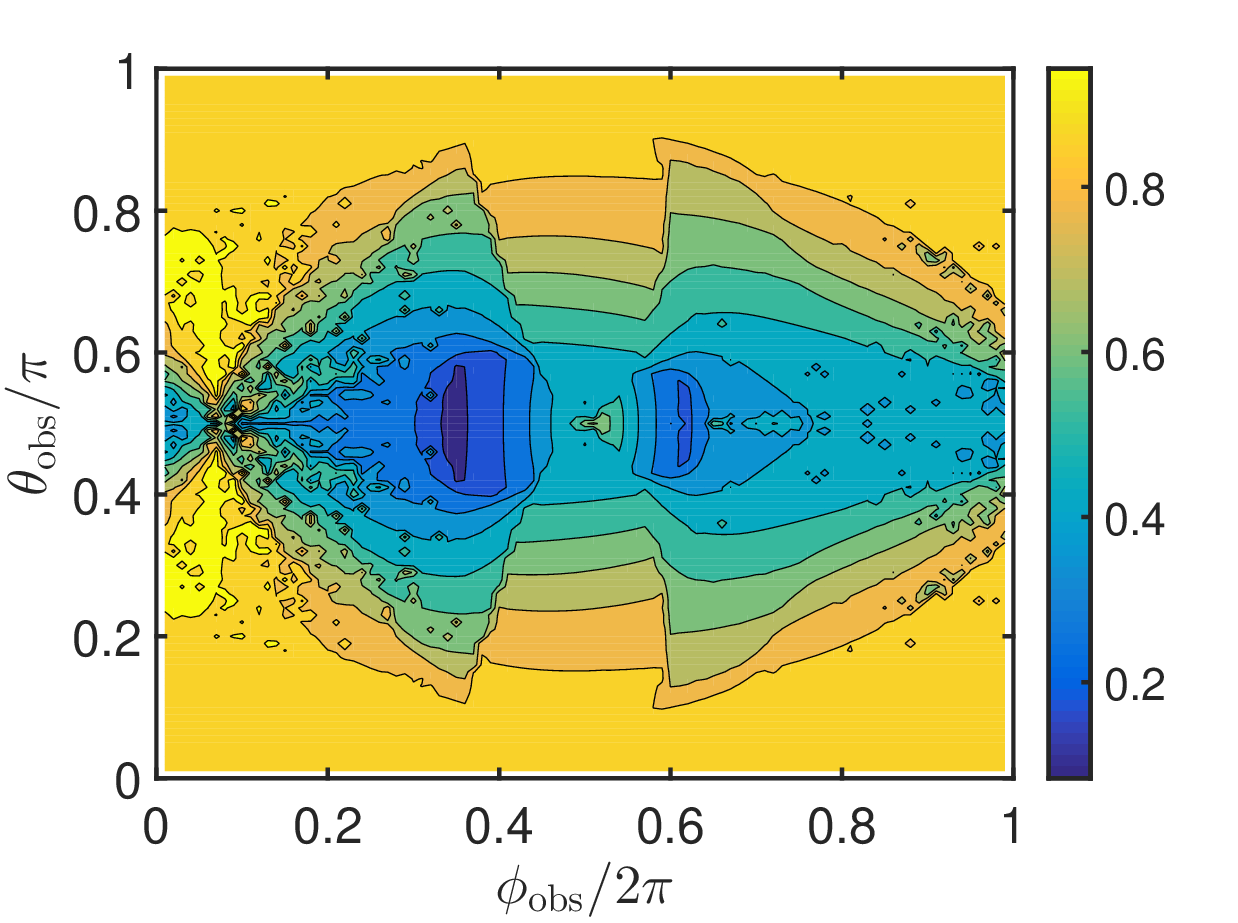}}
  \subfigure[]
  {\label{fig:f1d}
    \includegraphics[scale=.33]{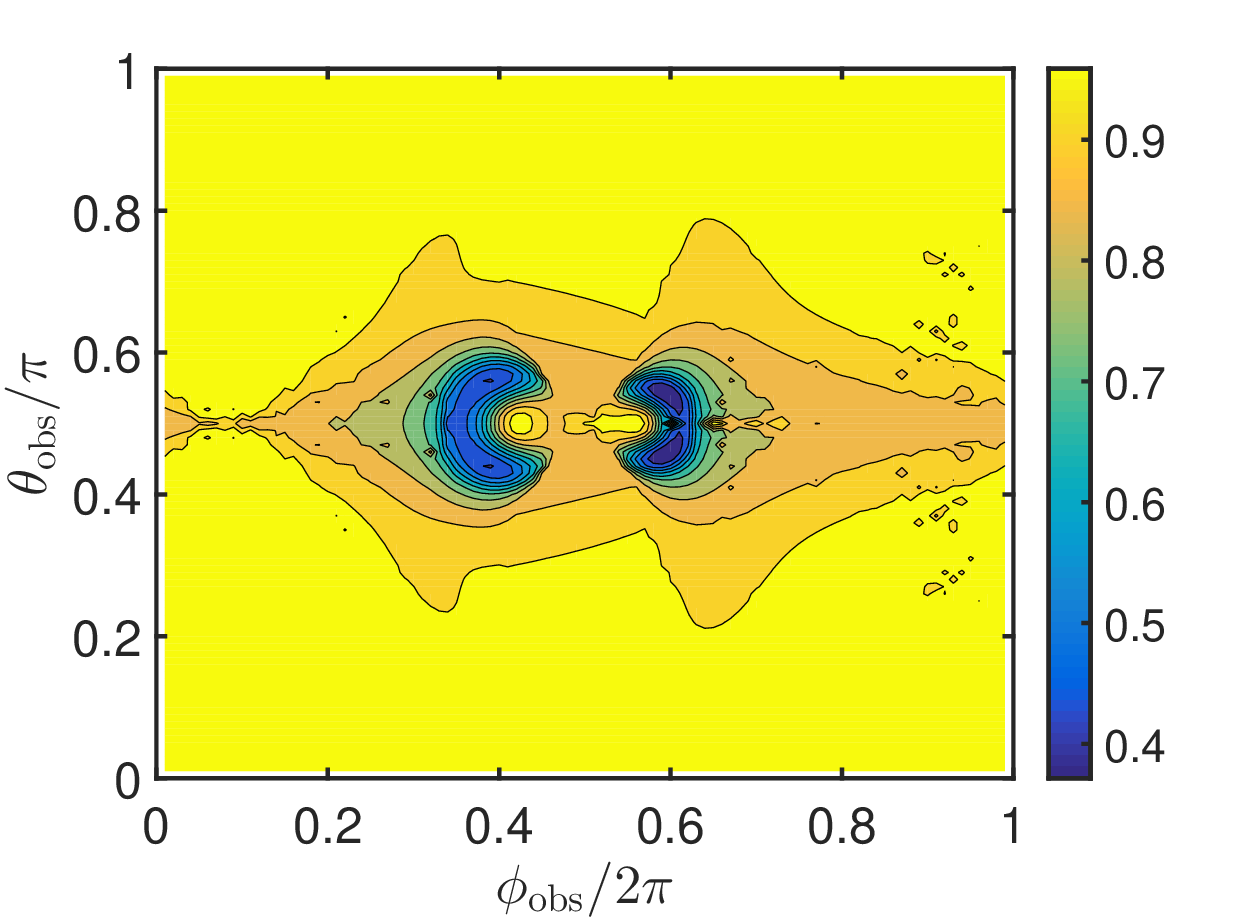}}
   \\
  \subfigure[]
  {\label{fig:f1e}
  \includegraphics[scale=.33]{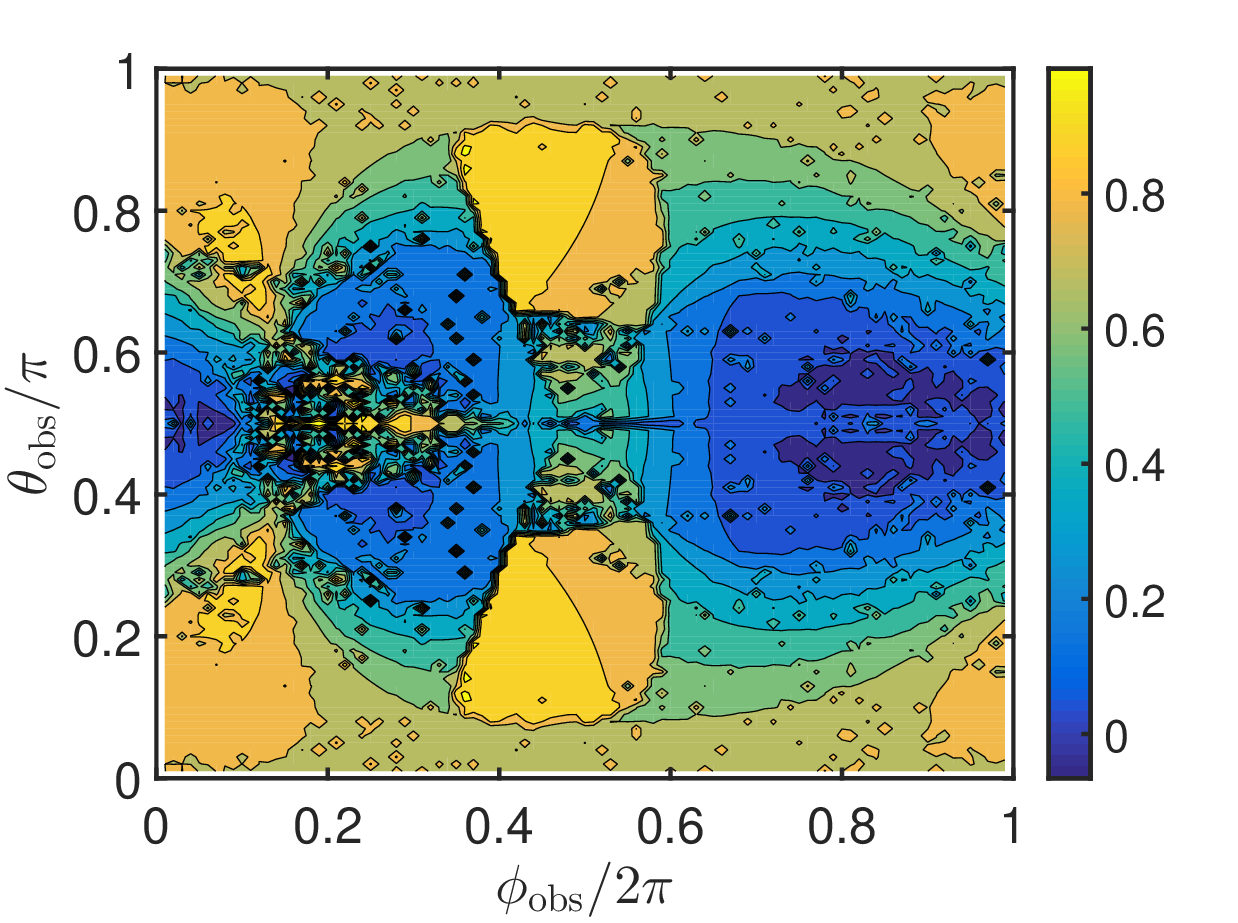}}
  \subfigure[]
  {\label{fig:f1f}
  \includegraphics[scale=.33]{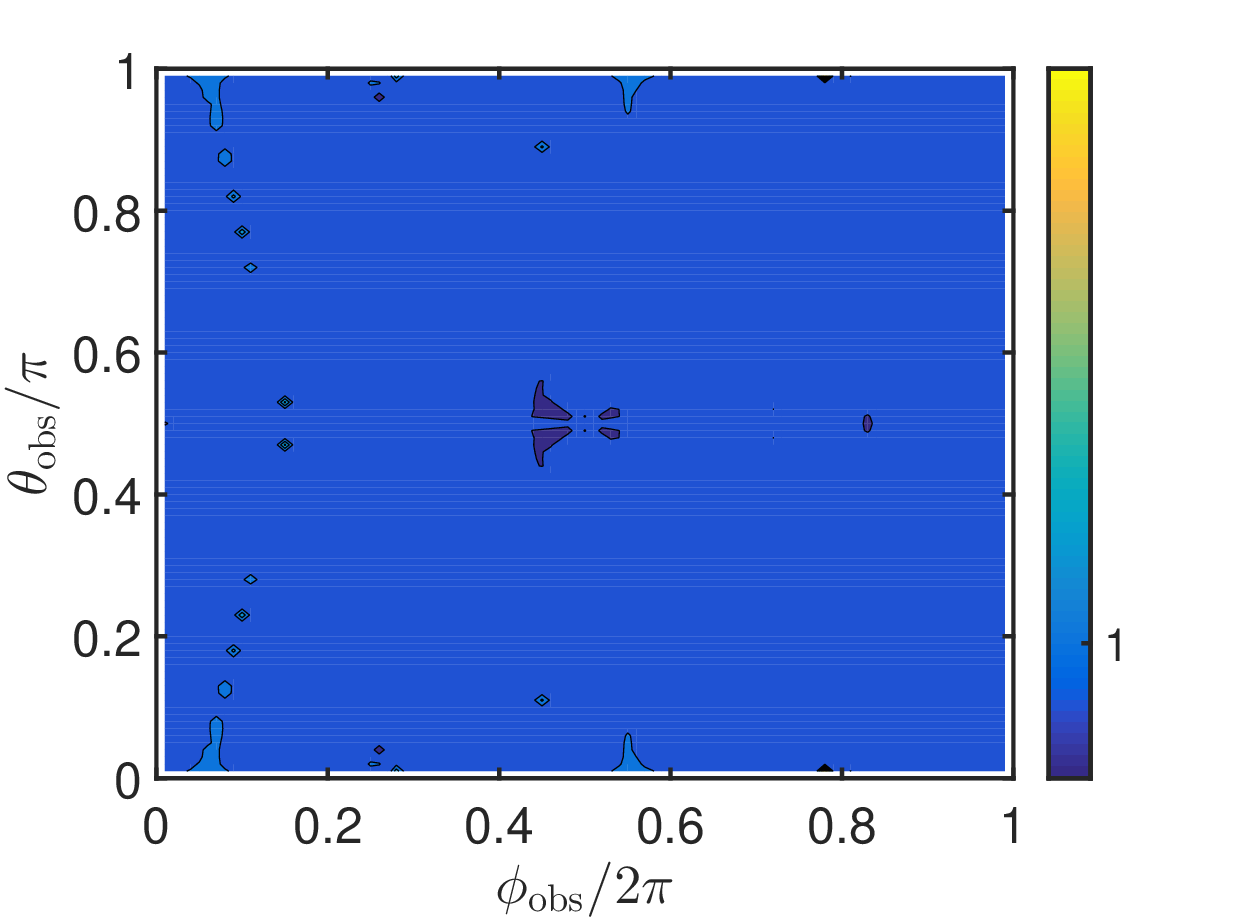}}
  \protect 
\caption{The fluxes of scattered neutrinos normalized by $F_{0}$ for SMBH with $M=10^{8}M_{\odot}$ and various BH spins. The maximal number
density in the disk is $10^{18}\,\text{cm}^{-3}$. The maximal strengths
of toroidal and poloidal fields are $320\,\text{G}$. The neutrino
magnetic moment is $10^{-13}\mu_{\mathrm{B}}$. Panels~(a) and~(b):
$a=2\times10^{-2}M$ ($z=10^{-2}$); panels~(c) and~(d): $a=0.5M$
($z=0.25$); ; panels~(e) and~(f): $a=0.98M$
($z=0.49$). Panels~(a), (c) and~(e) correspond to the poloidal field in Eq.~(\ref{eq:Atphi});
panels~(b), (d) and~(f) -- in Eq.~(\ref{eq:Aphi}).\label{fig:fluxpol}}
\end{figure}
\begin{table}[htpb]
  \begin{center}\setlength{\tabcolsep}{8pt}
    \begin{tabular}{|c|c|c|c|c|c|}
      \hline\hline
      &&&&&\\
	  {\bf Fig.~\ref{fig:f1a}}
          &{\bf Fig.~\ref{fig:f1b}}
          &{\bf Fig.~\ref{fig:f1c}}
          &{\bf Fig.~\ref{fig:f1d}}
          &{\bf Fig.~\ref{fig:f1e}}
          & {\bf Fig.~\ref{fig:f1f}}\\
	  \hline
          &&&&&\\
          $1\, 110\, 000$
          & $1\, 110\, 000$
          & $1\, 146\, 000$
          & $1\, 146\, 000$
          & $1\, 360\, 000$
          & $1\, 360\, 000$\\
	  \hline\hline
    \end{tabular}
  \end{center}
    \caption{Number of neutrinos in each figure. The actual number considered is in fact twice of each number since we also consider neutrinos below the equatorial plane.}
  \label{tab:number_of_neutrino}
\end{table}

\section{Results}
\label{sec:RES}

Active neutrinos are left polarized particles. If the neutrino spin precesses in an external field, a particle becomes right-handed, which cannot be observed in a neutrino detector. Thus, the observed flux of neutrinos is $F_\nu = P_{\mathrm{LL}} F_0$, where $F_0$ is the flux of scalar particles which is determined only by their geodesics motion. That is why our main goal is to find $F_\nu/F_0$ for scattered neutrinos.

We study the neutrino scattering off SMBH with $M = 10^8 M_\odot$. The BH spin varies in a quite broad range, $2\times 10^{-2} M < a < 0.98 M$. SMBH is supposed to be surrounded by an accretion disk consisting of a hydrogen plasma. This disk rotates with a relativistic velocity~\cite{Kom06}. The maximal number density of electrons is $n_e^{(\mathrm{max})} = 10^{18}\,\text{cm}^{-3}$~\cite{Jia19}. The disk is supposed to have a toroidal component of the magnetic field, which is inherent in the chosen accretion model. We assume that a nonzero poloidal magnetic field is present in the disk. We consider two models of such a field corresponding to Eqs.~\eqref{eq:Atphi} and~\eqref{eq:Aphi}. The maximal strength of both toroidal and poloidal field is $320\,\text{G}$, which is $\sim 1\%$ of the Eddington limit for this SMBH~\cite{Bes10}.

The electroweak interaction of neutrinos with the plasma of the accretion disk is within the forward scattering approximation~\cite{DvoStu02}. Neutrinos are supposed to be Dirac particles having a nonzero magnetic moment $\mu = 10^{-13}\,\mu_\mathrm{B}$, where $\mu_\mathrm{B}$ is the Bohr magneton. Such a neutrino magnetic moment is not excluded by observations of the globular clusters~\cite{Via13}. We consider neutrino spin oscillations within one neutrino generation. Thus, no transition magnetic moments are assumed.

The main disadvantage of the results of Refs.~\cite{Dvo23c,Dvo23d,Dvo23a,Dvo23b}, where the analogous problem has been studied, is the insufficient number of test particles in simulations. In this work, we overcome this obstacle by using parallel computing. We use Matlab for our numerical simulations. Since we can treat each neutrino independently, the loop over neutrino number is parallelized. We use both SkyLake and IceLake processors for our work. SkyLake has $24$ and IceLake has $38$ physical cores with large enough memory. The number of neutrinos we use in each case of BH spin and poloidal magnetic field is more than $2$ million (See Table~\ref{tab:number_of_neutrino}). To complete simulations for each of the cases takes a few days.

Note that only the gravitational interaction does not cause the change of the polarization of scattered ultrarelativistic neutrinos. The toroidal magnetic field in frames of the chosen model~\cite{Kom06} does not result in the neutrino spin flip either. These features have been established in Ref.~\cite{Dvo23b}.

In Fig.~\ref{fig:fluxpol}, we show $F_\nu/F_0$ for different spins of BH and various models of the poloidal magnetic field. Qualitatively, these results are similar to the findings of Ref.~\cite{Dvo23b}. However, we can track in more details the distribution of the poloidal field in the vicinity of BH with help of neutrinos. It becomes possible because of the significant enhancement of the test particle numbers in our simulations. Moreover, now we avoid the appearance of white gaps in contour plots in the case of the great spin of BH; cf. Figs.~\ref{fig:f1c}-\ref{fig:f1f}. We can see in Fig.~\ref{fig:f1f} the nontrivial distribution of $P_{\mathrm{LL}}$ which was unreachable in Ref.~\cite{Dvo23b}. This result again is possible because of the involvement of great number of neutrinos.

\section{Conclusion}
\label{sec:CONCL}

In the present work, we study the propagation and spin oscillations of the ultrarelativistic neutrinos in the gravitational field of SMBH surrounded by a magnetized accretion disk. These neutrinos are gravitationally scattered by the BH and interact electroweakly with the rotating matter of the disk and magnetically with the toroidal and poloidal fields. The magnetic interaction is possible because of the assumption of a nonzero neutrino magnetic moment. 

Despite the fact that the flux of the incoming neutrinos is parallel to the equatorial plane, these particles are supposed to move both above and below the equatorial plane. The geodesic motion of the neutrinos is taken into account exactly. The neutrino interaction with the external fields causes the particle spin precession, which is also accounted for along each neutrino trajectory.

The main improvement of the present work in comparison to Refs.~\cite{Dvo23c,Dvo23d,Dvo23a,Dvo23b} is the involvement of a large number of test particles in numerical simulations. Due to the use of parallel computing in this work, it becomes possible to use a large number of test particles. We use parallel facilities of Matlab on SkyLake and IceLake processors. These processors have large enough physical cores and memory to complete each computation in a matter of days.

Despite our results qualitatively similar to those in Ref.~\cite{Dvo23b}, now we are able to plot the survival probability in much more details. Thus one can probe more precisely the distribution of magnetic field in the accretion disk which $P_{\mathrm{LL}}$ depends on.

The obtained results can be used in the neutrino tomography of magnetic fields in the vicinity of BHs. Scattered neutrinos, emitted, e.g., in a supernova explosion, can be observed by the existing or future neutrino telescopes (see, e.g., Ref.~\cite{Abu22}). Our previous attempts to tackle this problem suffered from the insufficient  accuracy of calculations, which has been mentioned in Refs.~\cite{Dvo23c,Dvo23d,Dvo23a,Dvo23b}. Now, this shortcoming has been overcome  with the help of High Performance Parallel Computing.

\begin{acknowledgments}
  All our numerical computations have been performed at Govorun super-cluster at
  Joint Institute for Nuclear Research, Dubna.
\end{acknowledgments}

\end{document}